\documentclass[%
aip,
amsmath,amssymb,
reprint,%
]{revtex4-1}

\usepackage{graphicx}% Include figure files
\usepackage{dcolumn}% Align table columns on decimal point
\usepackage{bm}% bold math
\usepackage{multirow}

\usepackage[utf8]{inputenc}
\usepackage[T1]{fontenc}
\usepackage{mathptmx}
\usepackage{etoolbox}
\usepackage[dvipsnames]{xcolor}
\usepackage{changes}
\definechangesauthor[name=Adrian, color=red]{AL}

\usepackage{amsmath}
\usepackage{esint}
\newcommand{\eq}[2]{\begin{equation} \label{eq:#1} #2 \end{equation}}
\newcommand{\pd}{\partial}

\newcommand{\interfoam}{{\sffamily interfoam}}

\makeatletter
\def\@email#1#2{%
	\endgroup
	\patchcmd{\titleblock@produce}
	{\frontmatter@RRAPformat}
	{\frontmatter@RRAPformat{\produce@RRAP{*#1\href{mailto:#2}{#2}}}\frontmatter@RRAPformat}
	{}{}
}%
\makeatother

\begin{document}

	\preprint{AIP/123-QED}
	
	\title[Physics-based nozzle design rules for high-frequency liquid metal jetting]{Physics-based nozzle design rules for high-frequency liquid metal jetting}
	% Force line breaks with \\
	\author{J. Seo}
	\email{jseo@khu.ac.kr}
	\affiliation{Palo Alto Research Center, 3333 Coyote Hill Road, Palo Alto, CA 94304, USA}
	\affiliation{Department of Mechanical Engineering, Kyung Hee University, 1732 Deogyeong-daero, Giheung, Yongin, Gyeonggi 17104, Republic of Korea}
	\author{C. Somarakis}
	\affiliation{Palo Alto Research Center, 3333 Coyote Hill Road, Palo Alto, CA 94304, USA}
	\author{S. Korneev}%
	\affiliation{Palo Alto Research Center, 3333 Coyote Hill Road, Palo Alto, CA 94304, USA}
	\author{M. Behandish}
	\affiliation{Palo Alto Research Center, 3333 Coyote Hill Road, Palo Alto, CA 94304, USA}
	\author{A.J. Lew}
	\affiliation{Palo Alto Research Center, 3333 Coyote Hill Road, Palo Alto, CA 94304, USA}
	\date{\today}

    \begin{abstract}
	We present physics-based nozzle design rules to achieve high-throughput and stable jetting in drop-on-demand liquid metal 3D printing. The design rules are based on scaling laws that capture the change of meniscus oscillation relaxation time with geometric characteristics of the nozzle’s inner profile. These characteristics include volume, cross-sectional area, and inner surface area of the nozzle. Using boundary layer theory for a simple geometry, we show that the meniscus settles faster when the ratio of inner surface area to volume is increased. High-fidelity multiphase flow simulations verify this scaling. We use these laws to explore several design concepts with parameterized classes of shapes that reduce the meniscus relaxation time while preserving desired droplet specs. Finally, we show that for various nozzle profile concepts, the optimal performance can be achieved by increasing the ratio of the circumferential surface area to the bulk volume to the extent that is allowable by manufacturing constraints.
\end{abstract}

    \maketitle
    
    \section{Introduction} \label{sec:intro}

Metal additive manufacturing (AM) is emerging as a viable alternative to traditional methods such as casting to make supply chains more resilient and cost-effective for small-batch, multi-variety, and spare parts \cite{Sukhotskiy2018}. Drop-on-demand (DoD) liquid metal jetting stands out due to its high deposition throughput, low porosity, use of off-the-shelf materials (aluminum wire feed), predictable material properties, and operational safety\cite{Simonelli2019}. The process is characterized by ejecting a sequence of droplets from a microfluidic nozzle attached to the end of a pump where the metal is molten and pushed down at frequencies of a few hundred Hz using mechanisms ranging from magnetohydrodynamics\cite{Sukhotskiy2018} to pneumatics\cite{Luo2012}. As every droplet is released, the meniscus (liquid-gas interface) at the tip of the nozzle oscillates due to the dynamic interplay between surface tension, the inertia of the fluid, and the imposed pressure at the inlet of the nozzle (Fig. \ref{fig:nozzle}). Due to viscous dissipation in the fluid, the oscillations are damped.

To enable fast and reliable/repeatable builds, it is important to produce droplets with consistent volume, shape, and velocity distributions, which, in turn, strongly depend on the rate at which the post-ejection energy in the liquid in the nozzle dissipates. One way to quantify this rate is by observing the decay of the oscillations of the meniscus membrane at the tip of the nozzle. After a brief initial transient, the amplitude of these oscillations decay exponentially with a characteristic \emph{relaxation time}. Ideally, the meniscus should settle before the next droplet starts forming. 

The most common nozzle concepts are axisymmetric due to their ease of manufacturing (e.g., via micro-drilling). The resulting circular cross-sections lead to an inevitable tradeoff between droplet parameters, target throughput, and relaxation time. A nozzle with a smaller diameter has a smaller relaxation time, but it is harder to push the liquid through it  due to the need for larger fluid speeds to achieve a target throughput and the increased viscous resistance. As a result, a higher pressure is required to meet the  target throughput. Reducing the diameter of the nozzle may also lead to either smaller droplets or, through the use of an increased flow velocity, to an elongated droplet that breaks apart after the ejection. The critical question is to design a nozzle that ejects a largely spherical droplet with a target velocity and mass while minimizing the relaxation time of post-ejection oscillations. 

Analysis of the meniscus damping rate has been known as a challenging problem in multiphase fluid mechanics \cite{Howell1999,Ting1995} due to the non-linearity of the physics and of the partial differential equations associated with multiphase flows. Accurate analysis and prediction of the decay rate after a droplet ejection requires the solution of the Navier-Stokes equations through computational fluid dynamics simulations for extended periods of time. In our study, we identified the sources of viscous dissipation (both physical and numerical) and considered flows driven by the oscillatory dynamics induced by the meniscus in which inertial effects form a thin boundary layer. Previous studies on meniscus damping control for 3D printing \cite{Stacheswicz2009} adopted the one-dimensional harmonic-oscillator model, having a damper to model the viscous dissipation. These studies assumed fully-developed viscous flows (Poiseuille solution), but this assumption is invalid for the types of flow induced in liquid metals, which often have very low viscosity and for which the dynamics induced by the meniscus does not allow the boundary layer to develop. To the best of our knowledge, our work is the first study on the damping of a nozzle's meniscus considering  transient inertial effects,  and on applying these results to  establish design rules.

To demonstrate the application of the design rules, we performed physics simulations of multiphase flows capturing the oscillatory dynamics of the  liquid-gas interface on several nozzle geometries. We used a multiphase flow solver in OpenFOAM\textsuperscript{\textregistered}, which employs the algebraic volume-of-fluid approach for tracking gas-liquid interfaces. Our simulation results validated our hypothesis on the formation of a thin boundary layer near the nozzle wall, and demonstrated a successful decrease of the relaxation time in nozzles with increased surface-area-to-volume ratio. 

The paper is organized as follows: In Section \ref{sec:2}, we present the problem definition including the geometry, quantities of interest, governing equations, and non-dimensional parameters. In Section \ref{sec:3}, we present an analytical investigation of physical mechanisms that determine the nozzle performance, apply it to a simple cylindrical nozzle, and extend it to more complex shapes. In Section \ref{sec:4}, we describe the simulation setup used in our study. In Section \ref{sec:5}, we validate the identified scaling laws applying numerical simulations to solve the dynamics for a few nozzle concepts. % In Section \ref{sec:6}, we use the obtained scaling laws as design rules to generate a few designs that can outperform axisymmetric profiles, thereby enabling faster and higher-quality 3D prints. 
Section \ref{sec:6} summarizes and concludes the current findings. Lastly, in Section \ref{sec:7}, we elaborate on the numerical method we adopted to solve the governing equations and on the validation of the multiphase flow solver. 
    \section{Problem Setup} \label{sec:2}

\subsection{Nozzle Geometry} \label{sec:2-1}

\begin{figure}
	\vspace{-6pt}
	\centering
	\includegraphics[width=0.47\textwidth, keepaspectratio]{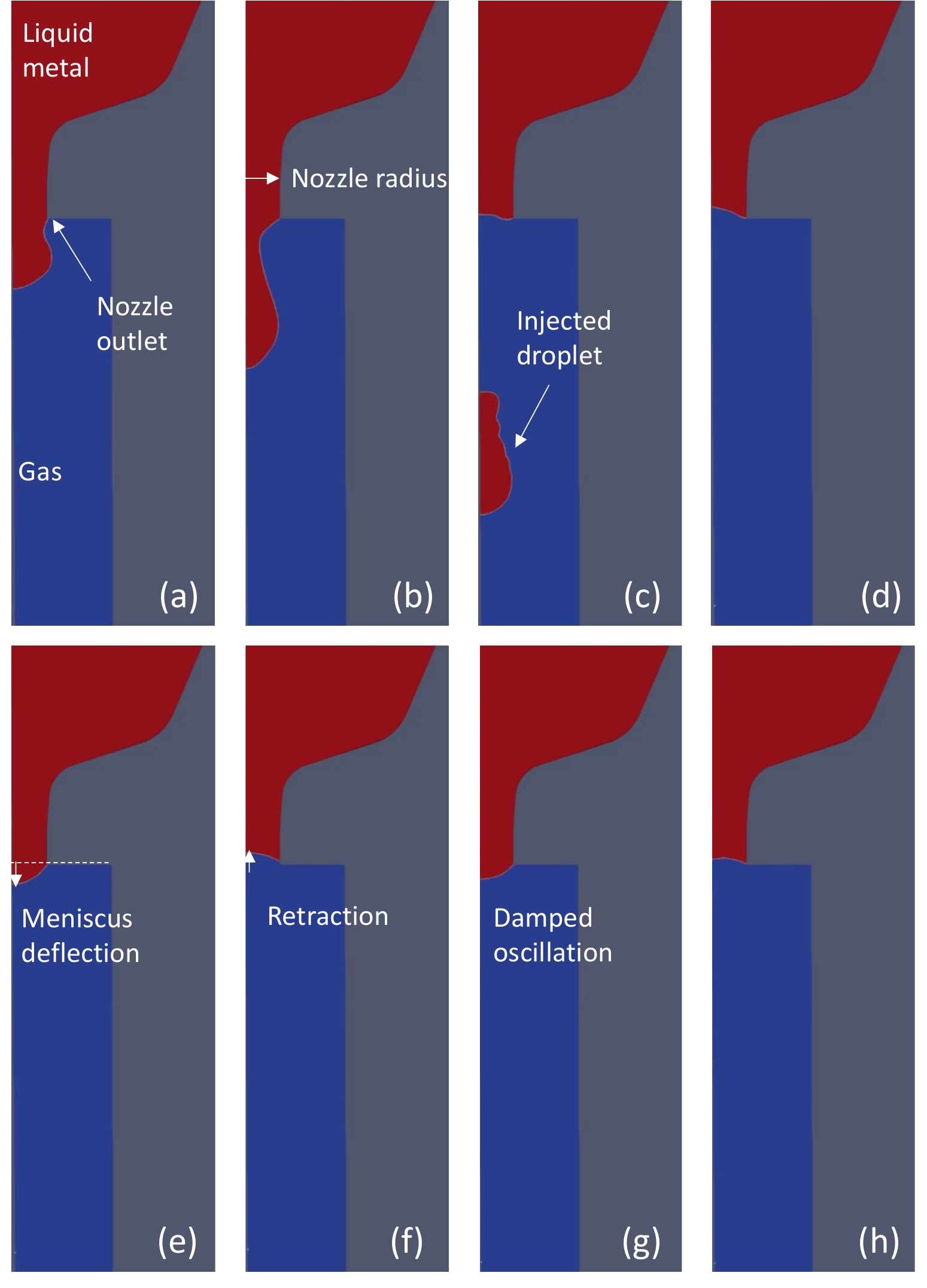}
	\caption{Time snapshots of the droplet ejection in the liquid metal jetting printer. The red colored area is filled with liquid metal, and the blue area corresponds to the area with gas. The  nozzle is axisymmetric, and is obtained by rotating the cross-section shown in the figures around the left edge, or the axis. (a-b) Droplet ejection by a high pressure pulse on the top surface; (c-d) meniscus retraction during the pressure pulsing; (e-h) damped oscillations of the meniscus without external pressure. Time intervals between snapshots vary.}
	\label{fig:nozzle}
\end{figure}

We consider a nozzle in a domain $V=A\times [0,L]$, where $A \subset \mathbb{R}^2$ is an open planar set that defines the  cross-section of the nozzle, and $L$ is its length. For $A$, we adopt either Cartesian coordinates $(x,y)$ or cylindrical coordinates $(r,\theta)$. The coordinate along $[0,L]$ will be denoted by $z$. The nozzle is filled with a viscous fluid with density $\rho$, dynamic viscosity $\mu$, and kinematic viscosity $\nu=\mu/\rho$. We assume that the dynamics in the nozzle starts with the fluid at rest at time $t=0$, and that the next droplet is ejected at time $t=T$. The nozzle outlet radius and diameter are denoted by $R$ and $d$, respectively. After a droplet ejection, surface tension drives the meniscus fluctuations and viscous dissipation dampens its deformation. The vertical meniscus displacement in the $z$ direction  is denoted by $\eta(x,y,t)$, and it is measured from the planar plane at the nozzle outlet ($z=0$). The contact line, where the liquid-gas-solid phases meet, is pinned at the sharp nozzle outlet, so  $\eta=0$ at $r=R$, as shown in Fig. \ref{fig:nozzle} (d). 

\subsection{Quantities of Interest} \label{sec:2-2}

Our goal when designing a nozzle for high-speed jetting is to reduce the relaxation time of the meniscus while retaining the ability to  eject droplets with a desired geometry and speed. We will describe the long-time dynamic behavior of the meniscus deformation, $\eta$, with that of a damped harmonic oscillator model  
\eq{eta}{\eta(x,y,t)=\eta_{t=0}(x,y) e^{-\gamma t}\cos(\omega t+\phi),}
where ${\eta_{t=0}}$ is the initial meniscus displacement, $\gamma>0$ is the damping rate,   $\omega$ is the oscillation frequency of the meniscus and $\phi$ is a phase. The relaxation time is $1/\gamma$. Shortly after a droplet is ejected, the dynamics of the meniscus is non-linear and cannot be fully described by a single mode damped harmonic oscillator model. After the initial transient, in which high wavenumber modes decay fast, we find that the long-time behavior of the meniscus can be well captured by the damped harmonic oscillator model with a single wavenumber mode. In the end, we aim to find a nozzle design geometry for which $\gamma$ is large enough to damp the oscillations within the ejection period (e.g., $e^{-\gamma T}\sim 10^{-2}$). In the long-time behavior, the value of $\gamma$ appears to be a feature of the geometry and the fluid only, and independent of the specific way in which a droplet is generated. 

\subsection{Governing Equations} \label{sec:2-3}

The dynamics of liquid and gas in and around the nozzle are governed by the Navier-Stokes (N-S) equation,
\begin{equation}
	\begin{split}
		\frac{\pd \rho_f \vec{u}}{\pd t} + \nabla \cdot (\rho_f \vec{u} \otimes \vec{u})=-\nabla p +\nabla\cdot  \mu_f(\nabla \vec{u}+\nabla \vec{u}^T) \\
		+\rho_f \vec{g} + \sigma \kappa\vec{n}\hat{\delta}(\vec{x}-\vec{x}_s) ,
	\end{split}
	\label{eq:NS}
\end{equation}
where $\vec{u}$ is the velocity field, $p$ is pressure, $\vec{g}$ is the acceleration of gravity, and $\sigma$ is the surface tension coefficient. $\rho_f$ is density of fluid either being $\rho$ for liquid metal and $\rho_g$ for gases. $\mu_f$ is viscosity of fluid, either being $\mu$ for liquid metal and $\mu_g$ for gases. The locus of the liquid-gas interface is described by a surface $\vec x_s$, $\kappa$ is its mean curvature, $\vec{n}$ is the local surface normal vector, and $\hat{\delta}(\vec{x}-\vec{x}_s)$ is the three-dimensional Dirac delta function. The last term in the right hand side of the equation represents the surface tension force, which is a concentrated force and non-zero only at the interface.  

We consider the algebraic Volume-of-Fluid (VoF) framework and introduce a three-dimensional scalar field $\alpha(x,y,z,t)$, defined as the volume fraction of liquid ranging from 0 to 1, to track the interface location $\vec{x}_s$. The (diffuse) interface between liquid and gas is algebraically described by setting $\alpha=0$ in the spatial domain filled with gas, and $\alpha=1$ in the spatial domain filled with liquid. In this way, the interface curvature and the surface normal vector can be calculated by  relations such as $\kappa =-\nabla  \cdot \vec{n}$ and  $\vec{n}=\nabla\alpha/|\nabla \alpha|$. The time evolution of the scalar field $\alpha$, and hence of the gas-liquid interface, is governed by the phase transport equation 
\eq{VOF}{\frac{\pd \alpha}{\pd t}+ \nabla \cdot(\vec{u} \alpha)=0.}

Finally, the system of equations is completed by stating the equation for mass conservation \eq{Continuity}{\nabla \cdot \vec{u} = 0,} at all points of the domain. This states the incompressibility of the liquid metal, and for simplicity, of the surrounding gas. 

We assume that the contact line is effectively pinned at the nozzle outlet rim. The nozzle outlet is macroscopically sharp, but assumed to be rounded at the microscale, so that the slope of the interface is not limited by the contact angle imposed by the material properties \cite{Seo2015}. With the fixed contact line, we assume negligible contribution from the contact line to the dissipation of energy in the system\cite{Snoeijer2013}. Our high-speed video recording of the meniscus (not shown here) confirmed that the meniscus can be considered to be pinned at the nozzle outlet. 

\subsection{Dimensional Analysis} \label{sec:2-4}

We proceed to derive  dimensionless numbers  to estimate the relative importance of different physical phenomena based on orders of magnitude approximations to typical material properties for a liquid metal and a gas. 

Typical values of mass density of liquid metals considered in  metal AM is $\rho\sim O(10^3)$kg/m$^3$, and the kinematic viscosity is $\nu \sim O(10^{-7})\text{m}^2/\text{s}$ (e.g. aluminum alloy) \cite{Assael2006}. The density of the surrounding gas (e.g. argon) is $\rho_g\sim O(1)$kg/m$^3$, and the kinematic viscosity is $\nu_g \sim O(10^{-6})\text{m}^2/\text{s}$. Surface tension at the interface between aluminum alloy\cite{Molina2007} and the argon gas is $\sigma \sim O(1)$N/m. We consider the range of nozzle diameter in the order of hundred micrometers $d\sim O(10^{-4})$m. The droplet size is mainly determined by the nozzle outlet diameter, so that the diameter of the droplet is $d_\text{Droplet}\sim d$. 

The droplet ejection rate for the droplet-on-demand liquid metal jetting is assumed  $f_\text{Jet} \sim O(10^2)$Hz. Based on the capillary flow theory \cite{Landau1958}, the meniscus oscillation frequency is $\omega \sim \sqrt{\sigma k^3/\rho}\sim O(10^3)Hz$, where $k$ is the wavenumber as $k\sim 1/d$ with $d= 500\mu m$. Therefore, we expect several periods of meniscus oscillations in between droplet ejections. 

Based on these values, we find that the Reynolds number associated with the oscillatory dynamics of the meniscus is $Re=\omega d^2 /\nu \gg O(1)$. Therefore, we expect the flow inside nozzle to involve inertial effects forming a thin boundary layer near the nozzle wall. A rough estimate of boundary layer thickness resulting from oscillating flow is $\delta \sim \sqrt{\nu/\omega}\sim O(10)\mu m$, which is thinner than the nozzle radius, $\delta/R\ll O(1)$. The flow inside the nozzle cannot be regarded as fully developed. 

The micro-scale nozzle size leads to capillary-force-dominant meniscus dynamics, in which the capillary length $l_c=\sqrt{\sigma/\Delta \rho g}=O(10^{-3})m$ is comparable or larger than the nozzle diameter, where $\Delta \rho$ is the difference in densities in the two fluids. Likewise, the Bond number is small, $Bo=gd^2\Delta \rho/\sigma \ll O(1)$, therefore the relative importance of the gravitational force on the meniscus deformation is negligible. The Weber number is $We=\rho U^2 d/\sigma \sim O(1)$, for $U\sim \omega d$ a characteristic speed, and the ejected droplet shape is determined by both inertial effect and the surface tension force. 

In sum, the dynamics of meniscus oscillation will strongly depend on the interplay between surface tension, inertia, and viscous forces in the liquid metal. We use the boundary layer theory for oscillatory flows \cite{Lamb1932,Landau1958} to analyze the flow physics. 
    \section{Physics-based Design Rules for Optimal Nozzle Performance} \label{sec:3}

In this section we present an analytical study of the physical mechanisms that determine the nozzle performance, and their scaling relations to the geometric parameters of the nozzle design. After deriving an analytical expression for the viscous dissipation from the Navier-Stokes equation, we develop scaling laws for the relaxation time $1/\gamma$ (alternatively, damping rate $\gamma$) for the cylindrical nozzle geometry, and introduce modified geometries with faster relaxation times based on these  law. 

\subsection{Source of Viscous Dissipation} \label{sec:3-1}

From the dimensional analysis in Section \ref{sec:2-4}, we assume that a thin viscous boundary layer is formed near the nozzle wall responding to the oscillatory fluid motion after the ejection of a droplet. A standard approach in potential flow analyses of classical fluid mechanics on multiphase flows \cite{Landau1958, Lamb1932} is to decompose the total damping rate $\gamma$ into two components. One source of dissipation is from the boundary layer, where the velocity gradient is dominated by the no-slip effect of the wall. Another source of viscous dissipation is from the bulk of fluid away from the wall, where the velocity gradient mainly depends on the flow structure. This approach has been extensively used to estimate the dissipation rate of sloshing liquid in containers \cite{Henderson1994, Case1956, Howell1999}. Accordingly, we decompose the total damping rate of the meniscus into two components, \eq{gamma_total}{\gamma=\gamma_\text{bulk}+\gamma_\text{bl},} where $\gamma_\text{bulk}$ is the damping rate from the bulk  of the  fluid and $\gamma_\text{bl}$ is the damping rate from the boundary layer. These damping rates stem from the contribution in each region to the dissipation of the kinetic energy of the fluid. To estimate them, we adopt standard flow structures (velocity fields) for the bulk of the fluid and the boundary layer. 

\subsection{Scaling of Damping Rate} \label{sec:3-2}

The damping rate can be estimated by the energy dissipation rate over the total energy of the system\cite{Landau1958}, \eq{gamma}{\gamma\sim \frac{\overline{\epsilon}}{\overline{E}},} where $\overline{\epsilon}$ is the mean (in time) viscous dissipation rate and $\overline{E}$ is the mean (in time) total energy in the system. Multiplying  the N-S equation (\ref{eq:NS}) by $\frac{1}{2}\vec{u}$ and integrating over the volume (i.e. kinetic energy form), we can derive the energy dissipation rate at a given time in a viscous fluid as \eq{diss}{\epsilon=\frac{\mu}{2} \iiint \nabla^s\vec u\colon\nabla^s\vec u \;dV,} where $\nabla^S\vec u$ is the symmetric gradient of the velocity field, and $A\colon B$ is the Frobenius inner product between tensors $A$ and $B$. The total energy of the system is \eq{energy}{{E}={\rho} \iiint |\vec{u}|^2 dV,} which is twice the kinetic energy \cite{Lamb1932}. The mean values of the dissipation rate and the total energy can be computed via time integration of the instantaneous values over a sufficiently long period of time. 

We first evaluate the scaling of the dissipation rate in the  bulk of the fluid, away from the wall. Following Eq. (\ref{eq:diss}), the scaling of the dissipation rate corresponding to a perturbed flow structure at a length scale $\lambda$ is \cite{Landau1958} \eq{eps_bulk}{\epsilon \sim \mu \frac{U^2}{{\lambda^2}} V,} where $U$ is the characteristic speed in the system, in our case, a characteristic speed of the meniscus, and $V$ the volume of the fluid. Given that the mean energy scales as \eq{e_bulk}{E \sim \rho U^2 V,} the damping rate in the bulk of the fluid scales \eq{gamma_bulk}{\gamma_\text{bulk}\sim \frac{\nu}{\lambda^2}. } This indicates that perturbed flow structures with smaller size will decay faster. Our major focus will be on the largest wavenumber mode, which stays in the system for the longest time. 

For the boundary layer dissipation near the wall, the surface integral form of Eq. (\ref{eq:diss}) is considered, \eq{eps_bl}{\epsilon \sim \mu \frac{U^2}{\delta} S,} where $S$ is the surface area exposed to solid surface and $\delta$ is the boundary layer thickness formed near the solid surface. Under the oscillatory flow with frequency $\omega$, a dimensional analysis on the boundary layer thickness $\delta$ estimates \cite{Landau1958} \eq{BL}{\delta\sim \sqrt{\frac{\nu}{\omega}}.} Finally, the damping rate at the boundary layer of the fluid scales as \eq{gamma_bl}{\gamma_\text{bl} \sim \sqrt{\nu\omega}\frac{S}{V},} where Eq. (\ref{eq:energy}) is used to estimate the kinetic energy. The oscillation frequency of the meniscus can be roughly estimated as $\omega \sim \sqrt{\sigma /\rho R^3}$. The scaling relation Eq. (\ref{eq:gamma_bl}) indicates that the interface deformation decays faster when the surface area to the volume ratio is large. This provides an important insight on the nozzle design and we will use this rule to control the dissipation rate. 

Potential flow analyses \cite{Landau1958,Case1956} identified the exact relation in Eq. (\ref{eq:gamma_bulk}) for ideal waves as $\gamma_\text{bulk}=2\nu k^2$, where $k$ is the wavenumber of the flow structure of interest. For example, in a closed brimful cylinder, the lowest wavenumber is $k=\beta/R$, where $\beta\simeq 3.83$ is the first root of the derivative of the first Bessel function of the first kind\cite{Case1956}. Furthermore, the boundary layer dissipation rate identified by linear analyses \cite{Case1956} is $\gamma_\text{bl}\simeq\sqrt{\frac{\nu\omega}{2}}\frac{1}{2R}$, for a cylinder with radius $R$ and height $L$ ($V=\pi R^2 L$ and $S=2\pi R L$). We note that although this linear solution provided an accurate estimate of the decay rate in small amplitude capillary waves ($L\gg R$)\cite{Case1956,Henderson1994, Howell1999}, it does not guarantee accurate predictions when the meniscus dynamics is non-linear due to large interface fluctuations\cite{Ting1995}. However, we expect the damping rate to scale similarly with $S$ and $V$ even during the initial, non-linear stages of the meniscus oscillations.

Using the insights from the scaling, we draw the following nozzle design rules for high speed droplet jetting: 
\begin{itemize}
	\item Given  the inverse scaling of $\gamma_\text{bulk}$ and $\gamma_\text{bl}$ with the size of the system (with both $\lambda$ and $\frac{V}{S}$), both damping rates can be controlled by shaping (i.e. constricting) the cross-section  of the nozzle.
	\item Given that $\gamma_\text{bl} \sim S/V$, the damping rate can be controlled by maximizing the surface area of the nozzle. 
\end{itemize}
In the next subsections, we apply these ideas to estimate the expected scaling of the damping rate upon a  design change of the nozzle. 

\subsection{Dissipation Control by Nozzle Constriction} \label{sec:3-3}

\begin{figure}
	\vspace{-6pt}
	\centering
	\includegraphics[width=0.47\textwidth, keepaspectratio]{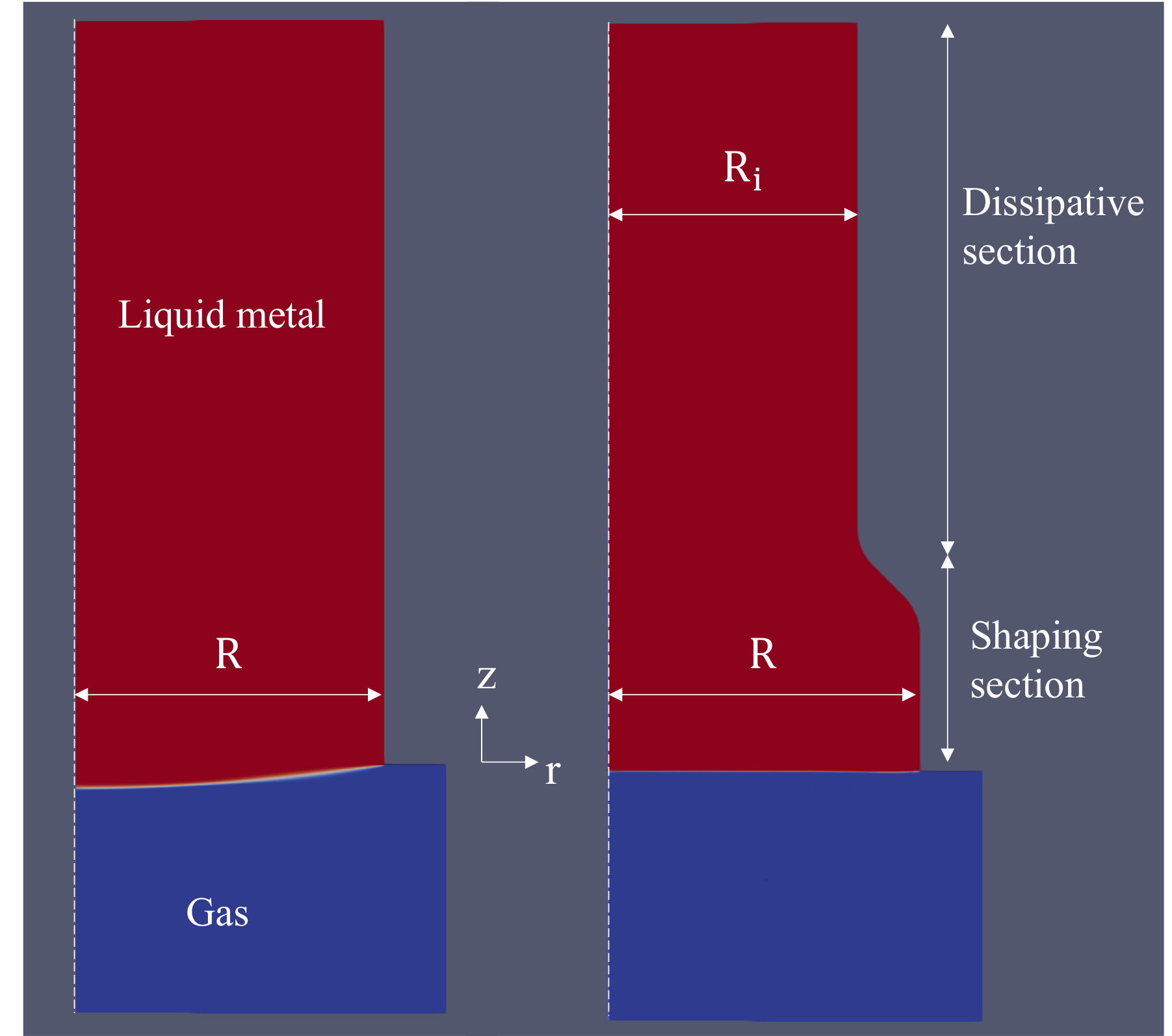}
	\caption{Geometric configurations of nozzles with constricted cylinders. The  nozzle is axisymmetric, and is obtained by rotating the cross-section shown in the figures around the left edge, or the axis, indicated by a dashed line. Here $R$ is the radius of the nozzle and $R_i$ is the radius of the constricted part. (Left) Standard cylindrical nozzle, showing a deformed liquid-gas interface; (right) constricted nozzle with $R_i$=$0.8R$ and a flat interface. The upper section with diameter $R_i$ is termed the {\it dissipative section}, while the lower section with diameter $R$ is referred to as the {\it shaping section}.}
	\label{fig:constricted}
\end{figure}

As expected, the scaling of the damping rate identifies the diameter of the nozzle as the parameter that most significantly defines the damping rate for a cylindrical nozzle. This naturally motivates the design of a  nozzle with a smaller diameter, for damping control. However, the diameter of the nozzle is limited by the targeted droplet size. To account for this constraint, we consider nozzles with two distinct sections: a constricted section destined to control the damping rate, separated from another section that controls the droplet shape. An example of a nozzle with these two sections is found in Fig. \ref{fig:constricted}. The part of the nozzle with a smaller diameter $R_i<R$ is  referred to as the {\it dissipative section}, whereas the section near the nozzle outlet with radius $R$, whose role is to control the the droplet size, is referred to as the {\it shaping section} (Fig. \ref{fig:constricted}). 

\subsection{Dissipation Control by Increasing Surface Area to Volume} \label{sec:3-4}

\begin{figure}
	\vspace{-6pt}
	\centering
	\includegraphics[width=0.48\textwidth, keepaspectratio]{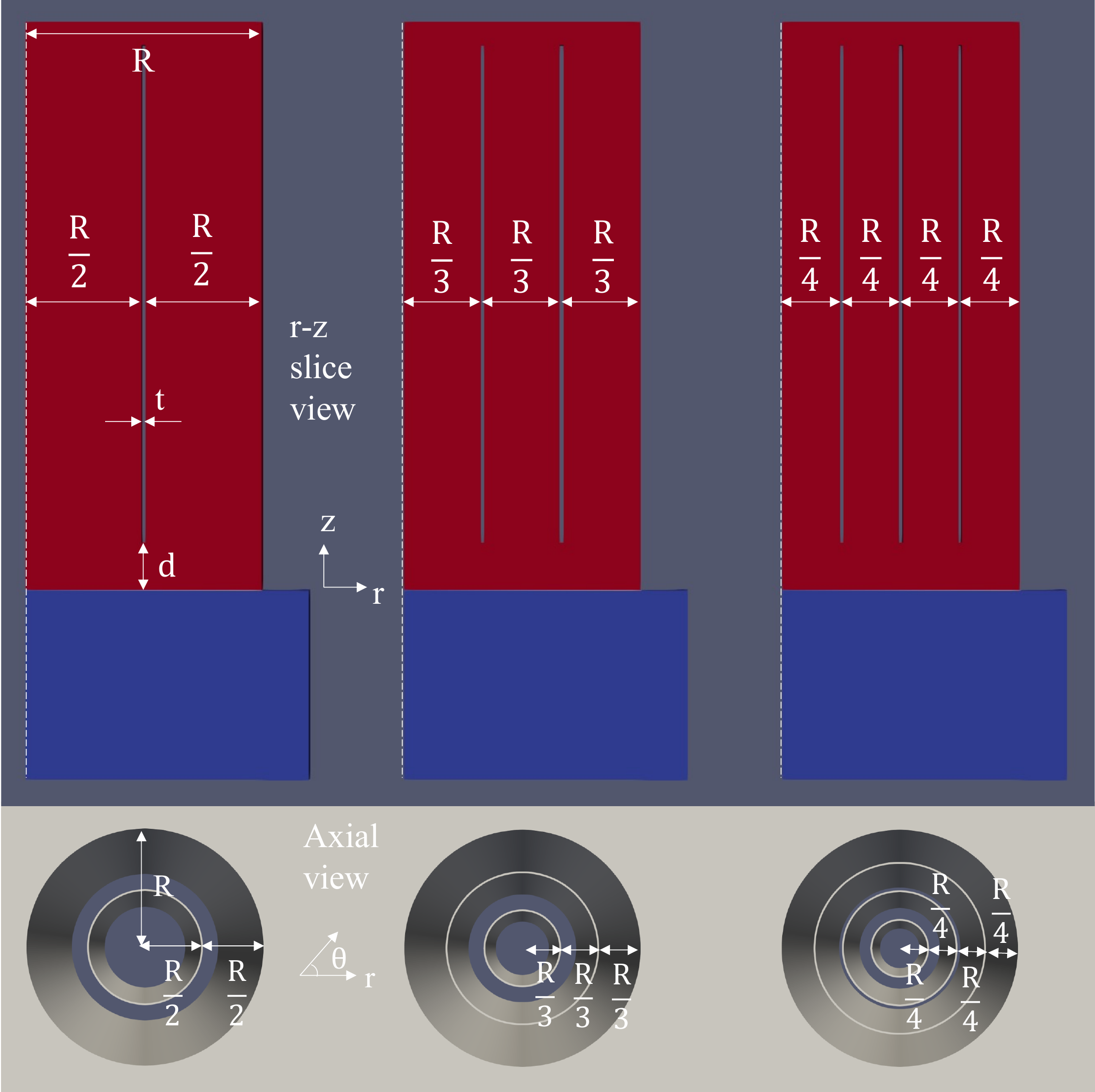}
	\caption{Geometric configurations of nozzles with concentric cylinders. (Top) vertical slice of the nozzles with concentric cylinders. The computational domain is designed for axisymmetric simulations. The inner cylinders are located above the nozzle outlet by a distance $d$. The thickness of the inner cylinders is denoted as $t$. All inner cylinders have the same thickeness. Dashed lines indicate axes of rotation. (Bottom) Horizontal slices of nozzles with concentric cylinders. Dark gray regions are the inner walls of the cylinder. From left to right, n=2,3,4.}\label{fig:cyls}
\end{figure}

Based on the scaling relation, Eq. (\ref{eq:gamma_bl}), the second method to reduce the relaxation time is to maximize the surface area to volume ratio ($S/V$). The main advantage of this method is that we can control the dissipation without changing the cross-sectional area, and hence can circumvent large changes of the mean speed of the fluid in the nozzle. 

To show the proof-of-concept of this design rule, we considered an (idealized) cylindrical nozzle with $n$ total concentric cylinders identically spaced in the radial direction, as shown in Fig. \ref{fig:cyls}. This idealized geometry is a great example of design that demonstrates larger surface area compared to a typical cylinder nozzle, while keeping the cross-sectional area essentially the same. We consider the thickness of the inner cylinders in the radial direction, $t$, much smaller than the radius of the outer cylinder, to keep the cross-sectional area largely constant. In this way, we control the dissipation while not changing the dynamics of the meniscus, which is mainly controlled by the outlet diameter. 

The total surface area of nozzle with $n\geq 1$ number of cylinders\footnote{Here $n=1$ is the standard nozzle, i.e. with zero internal concentric cylinders.} is $S_\text{n}=2\pi RL n$, therefore, the resulting dissipation rate scales \eq{gamma_bl_n}{\gamma_\text{bl,n}\sim \sqrt{\nu\omega} \frac{n}{R}.} We expect that the bulk dissipation will not change significantly if the axial component of the velocity is the dominant flow feature associated with the damping. 
    \section{Simulation Setup} \label{sec:4}

To demonstrate the effectiveness of our design rule, we employed the open-source CFD software OpenFOAM\textsuperscript{\textregistered} and conducted multiphase flow simulations of the meniscus dynamics in the nozzle with several different designs. We studied the relaxation time of the meniscus suspending at the outlet of the nozzle responding to a pressure pulse at the nozzle inlet. The meniscus was initially flat and  pinned at the nozzle outlet.

We chose material properties and nozzle dimensions inspired by  3D printing scenarios. We set the material properties to those of a liquid aluminum alloy \cite{Assael2006} $\rho=2435 kg/m^3$, $\nu=4.16 \times 10^{-7}m^2/s$, and of argon gas $\rho_g =1.6kg/m^3$, $\nu_g=2.6\times 10^{-5}m^2/s$, and surface tension\cite{Molina2007} $\sigma=0.85N/m$. 

In this study, we considered $R=250\mu m$ as the nominal radius of the outlet of the nozzle. For the constricted nozzle (Fig. \ref{fig:constricted}), we set $R_i=200 \mu m$ and $R_i=150 \mu m$. The connecting area between the dissipative section and the shaping section had smooth curves with a finite curvature to minimize the formation of vortices at sharp corners, although flow separation can still be observed. For concentric cylinders, we placed equally spaced inner cylinders along the radial direction (Fig. \ref{fig:cyls}). The thickness of each inner cylinder is $t=4\mu m$. The inner cylinders were pushed inward by a distance $d=50\mu m$ so that the meniscus did not directly touch the inner cylinders during oscillations. This last part effectively acts as the shaping section of this nozzle.

The OpenFOAM thin wedge geometry was employed to represent planes in the swirl direction for 2-D rotationally symmetric cases. The axi-symmetric wedge geometries (shown in Fig. \ref{fig:constricted} and Fig. \ref{fig:cyls}) were spatially discretized into approximately 300K-400K hexaderal mesh elements (400 elements in the radial direction, 900 elements in the vertical direction inside the liquid domain). We used a non-uniform grid and applied extra grid refinement in the region  the liquid-gas interface sweeps during oscillations. Our multi-step grid refinement study confirmed the convergence of the quantities of interest. 

The simulations were performed until 4.5 millisecond after the initial transident time period. We conducted high-performance parallel computing simulations with 48 processors on the Amazon AWS clusters. 

\section{Proof of the Design Rule} \label{sec:5}

\subsection{Standard Cylindrical Nozzle} \label{sec:5-1}

\begin{figure}
	\vspace{-6pt}
	\centering
	\includegraphics[width=0.45\textwidth, keepaspectratio]{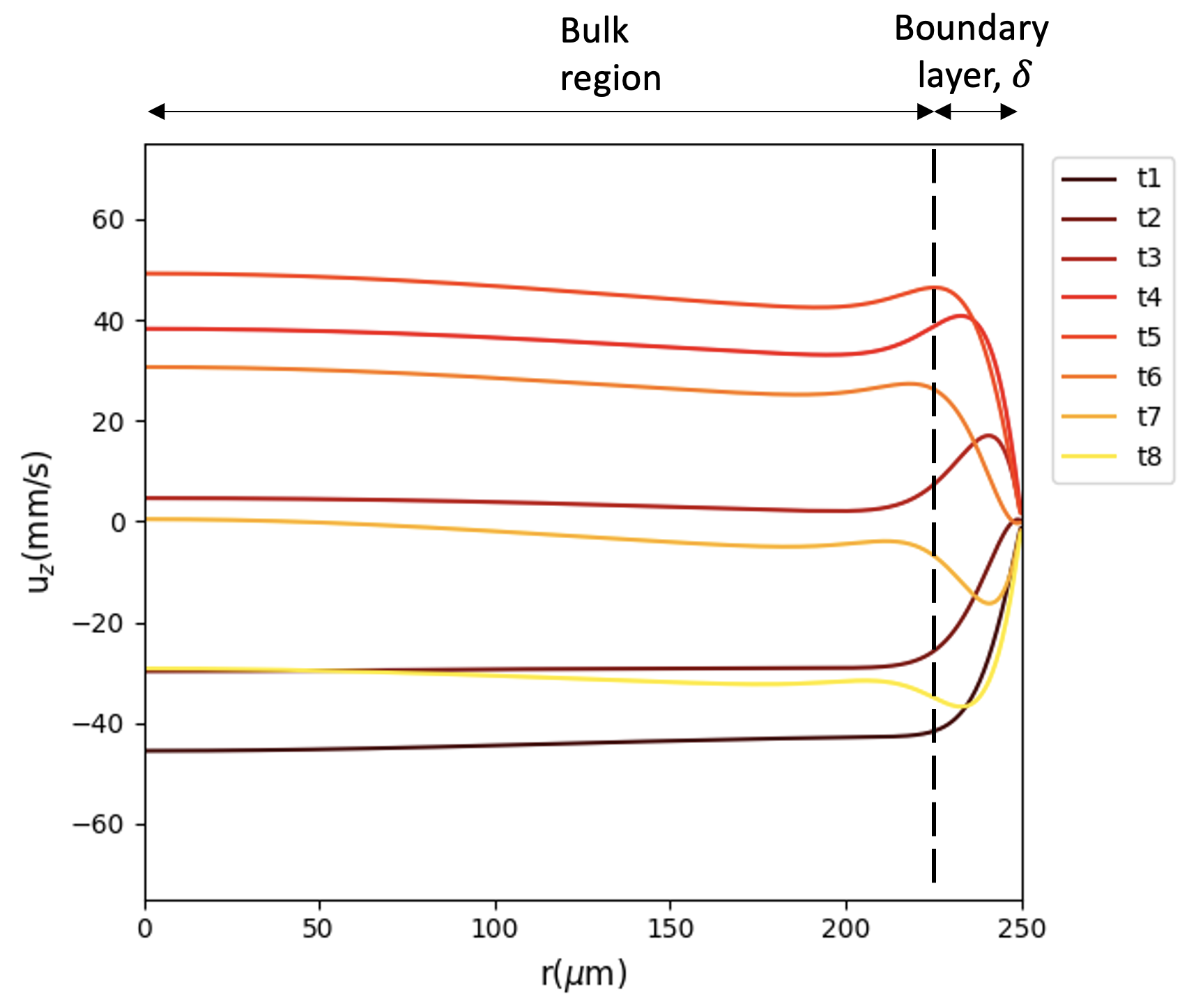}
	\caption{Time snapshots of velocity profiles generated by the meniscus oscillation inside the cylindrical nozzle after a pressure pulse is applied on top. The cylindrical nozzle radius is $R=250\mu m$. From brown to yellow ($t_1-t_8$), the vertical velocity $u_z$ is plotted in the increasing order in time in every 0.1 ms. The velocity profiles are measured
		at $z=200\mu m$ above the nozzle outlet.}
	\label{fig:bl}
\end{figure}

Figure \ref{fig:bl} shows instantaneous time snapshots of the vertical velocity profiles every 0.1 milliseconds at $z=200\mu m$ above the nozzle outlet. From the  velocity profiles near the nozzle wall, it is evident that our hypothesis that the meniscus dynamics forms a thin oscillatory boundary layer near the wall is valid. The steep velocity gradient near the wall is confined to the near wall region up to $R-\delta<r<R$ with $\delta \approx 20 \mu m$. The observed boundary layer thickness is consistent with the estimate from the oscillatory viscous flow theory, $\delta \sim \sqrt{\nu/\omega}\approx18 \mu m$. 

While the velocity gradient inside the boundary layer is governed by the no-slip condition on the wall, the bulk region of the fluid has almost flat velocity profiles with negligible velocity gradients. The plug-like oscillatory flow is driven by the motion of the meniscus.  This clear difference in velocity gradient in the bulk and the boundary layer indicates that the boundary layer dissipation is the major source of damping.

\begin{figure}
	\vspace{-6pt}
	\centering
	\includegraphics[width=0.5\textwidth, keepaspectratio]{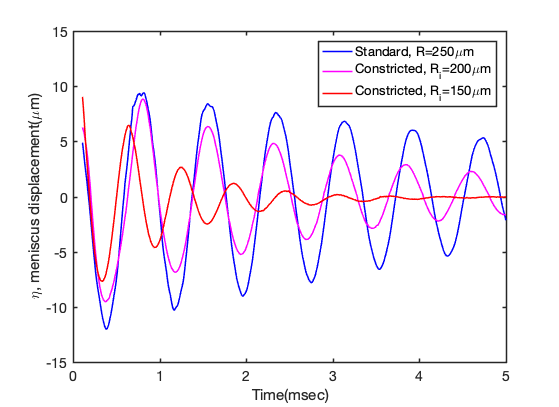}
	\caption{Time history of the meniscus displacement at the center of the channel, $\eta_0(t)=\eta(r=0,t)$, for the standard cylindrical nozzle with $R=250\mu m$ and the nozzle with a constricted section with radius $R_i=200\mu$ and $R_i=150\mu m$.}\label{fig:dis_cons}
\end{figure}

Our main quantity of interest is the damping rate of the amplitude of the meniscus motion, $\gamma$. We post-processed the time history of the interface location at the center of the nozzle, $\eta_0(t)=\eta(t,r=0)$, in Fig. \ref{fig:dis_cons}. At each time $t$, the interface apex location was obtained via fitting $[1+\tanh(z-\eta_0(t))]/2$ to the continuous volume fraction field along the axis, to find the value of $\eta_0(t)$. On the time history of the interface, we calculated $\gamma$ and $\omega$ by fitting a damped sinusoidal function $f(\gamma, \omega, \phi, A)=A \text{exp}(-\gamma t) \text{cos}(\omega t+\phi)$ using the curve fit tool provided in the {\sffamily Scipy.optimize}. To disregard initial transient effects, only data after 0.5 milliseconds were used in the fit. As a result, the decay rate of the meniscus for the cylindrical nozzle (shown in Fig. \ref{fig:constricted}) with $R=250\mu m$ is obtained as 

\eq{gamma_cyl}{\gamma =167/\text{sec},} and $\omega$=1269/sec. The variance associated with the fitted parameters are $\pm5$/sec for $\gamma$, and $\pm1$/sec for $\omega$ with 95$\%$ confidence interval. All 95$\%$ confidence bounds for $\gamma$ and $\omega$ in this paper, although not specified, have less than 2.6 percent of their reported values. 

\subsection{Nozzle with a Constriction} \label{sec:5-2}

The time evolution of the meniscus amplitude for standard and constricted nozzles is plotted in Fig. \ref{fig:dis_cons}. The nozzle with a constricted section successfully leads to faster dissipation of the meniscus oscillation. The decay rates of the meniscus displacement with the constricted nozzle are significantly larger than the standard nozzle as shown in Table \ref{table:1}. 

\begin{table}[ht]
	\begin{tabular}{ |c||c|c|c| } 
		\hline
		Nozzle type & Radius at constriction&$\gamma(\text{sec}^{-1})$ & $\omega(\text{sec}^{-1})$ \\
		\hline
		Standard &$250\mu m$ & 167 & 1269 \\ 
		Constricted &$200\mu m$& 358 & 1318 \\ 
		Constricted&$150\mu m$& 1176 & 1655\\ 
		\hline
	\end{tabular}
	\caption{Damping rate $\gamma$ and oscillation frequency $\omega$ of the meniscus motion for the standard nozzle and the nozzles with $n$ concentric cylinders. }
	\label{table:1}
\end{table}

This result demonstrates that only a fraction reduction of the nozzle radius (here 20$\%$-40$\%$) was effective to increase the decay rate by approximately factor of 2-7. The higher dissipation rate at more constricted nozzle can be explained by the trends from both $\gamma_\text{bl}$ and $\gamma_\text{bulk}$. The constricted cylinder has higher boundary dissipation rate due to larger surface area to the volume ratio as well as higher bulk dissipation rate due to smaller diameter against the standard cylindrical nozzle. The relationship between $\gamma$ and $R_i$ is non-linear due to combination of these two factors and the contribution from the intermediate section between the shaping and the dissipative section. The direct proportionality between the surface area to volume ratio and  $\gamma_\text{bl}$ is investigated via the concentric cylinders in the following section. 

\subsection{Nozzle with Concentric Cylinders} \label{sec:5-3}

\begin{figure}[ht]
	\vspace{-6pt}
	\centering
	\includegraphics[width=0.45\textwidth, keepaspectratio]{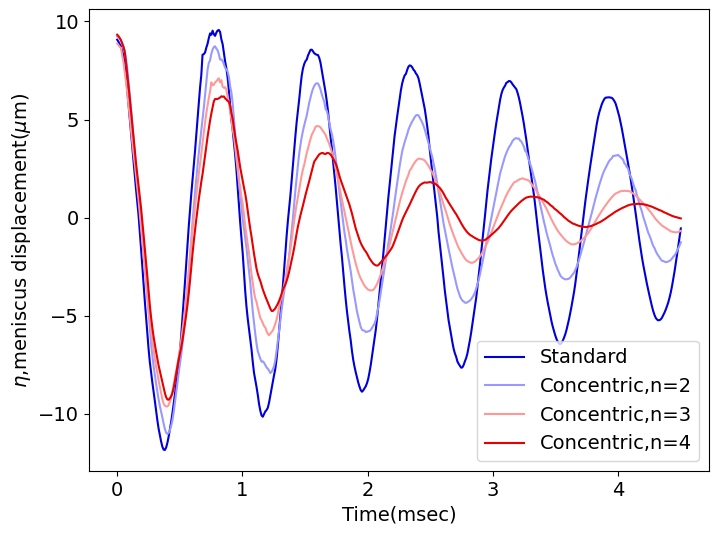}
	\caption{Time history of the meniscus displacement at the center of the channel, $\eta_0(t)=\eta(r=0,t)$, for the standard cylindrical nozzle with $R=250\mu m$ and a nozzle of the same radius with concentric inner cylinders. Here $n$ is the  number of concentric cylinders in the nozzle.}\label{fig:dis_conc}
\end{figure}

The time evolution of the meniscus amplitude for nozzles with different number of concentric  cylinders inside is plotted in Fig. \ref{fig:dis_conc}. The amplitude of the meniscus motion decays faster in nozzles with larger surface area, therefore the concentric cylinders are effective at  reducing the relaxation time.  We summarized the damping rate and the oscillation frequency for $n=1,2,3,4$ in  Table \ref{table:2}. The results show that the damping rate increases linearly (with a slope of $\approx 200$/sec) and confirms the proposed scaling of the damping rate dominated by the boundary layer, $\gamma_\text{bl}\sim n$ in Eq. (\ref{eq:gamma_bl_n}). Notice that the  oscillation frequency remains largely fixed, since the outlet diameter and total mass inside the nozzle are unchanged. 

\begin{table}[ht]
	\begin{tabular}{ |c||c|c| } 
		\hline
		Nozzle type & $\gamma(\text{sec}^{-1})$ & $\omega(\text{sec}^{-1})$ \\
		\hline
		Standard nozzle, n=1 & 167 & 1269 \\ 
		Concentric cylinders, n=2& 343 & 1256 \\ 
		Concentric cylinders, n=3& 550 & 1237\\ 
		Concentric cylinders, n=4& 768 &  1205\\ 
		\hline
	\end{tabular}
	\caption{Damping rate $\gamma$ and oscillation frequency $\omega$ of the meniscus motion for the standard nozzle and the nozzles with $n$ concentric cylinders. }
	\label{table:2}
\end{table}

\begin{figure}
	\vspace{-6pt}
	\centering
	\includegraphics[width=0.46\textwidth, keepaspectratio]{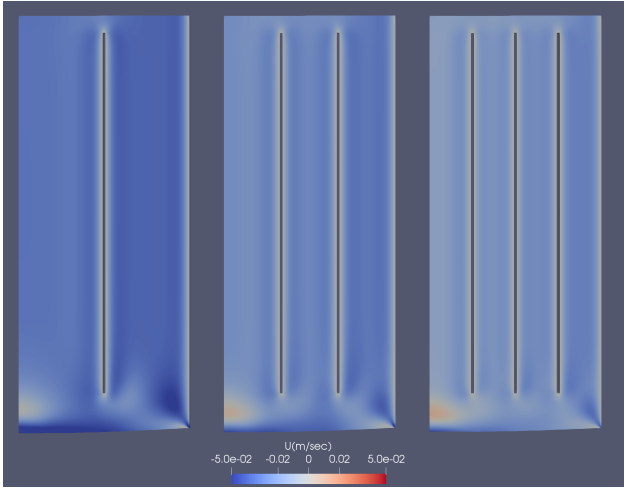}
	\caption{Instantaneous snapshots of the contours of the axial velocity component in the liquid domain, $u_z$, at $t=1$ msec for the nozzle with concentric inner cylinders. The countour plots show boundary layers near the solid wall surfaces.}\label{fig:velocity}
\end{figure}

\begin{figure}
	\vspace{-6pt}
	\centering
	\includegraphics[width=0.5\textwidth, keepaspectratio]{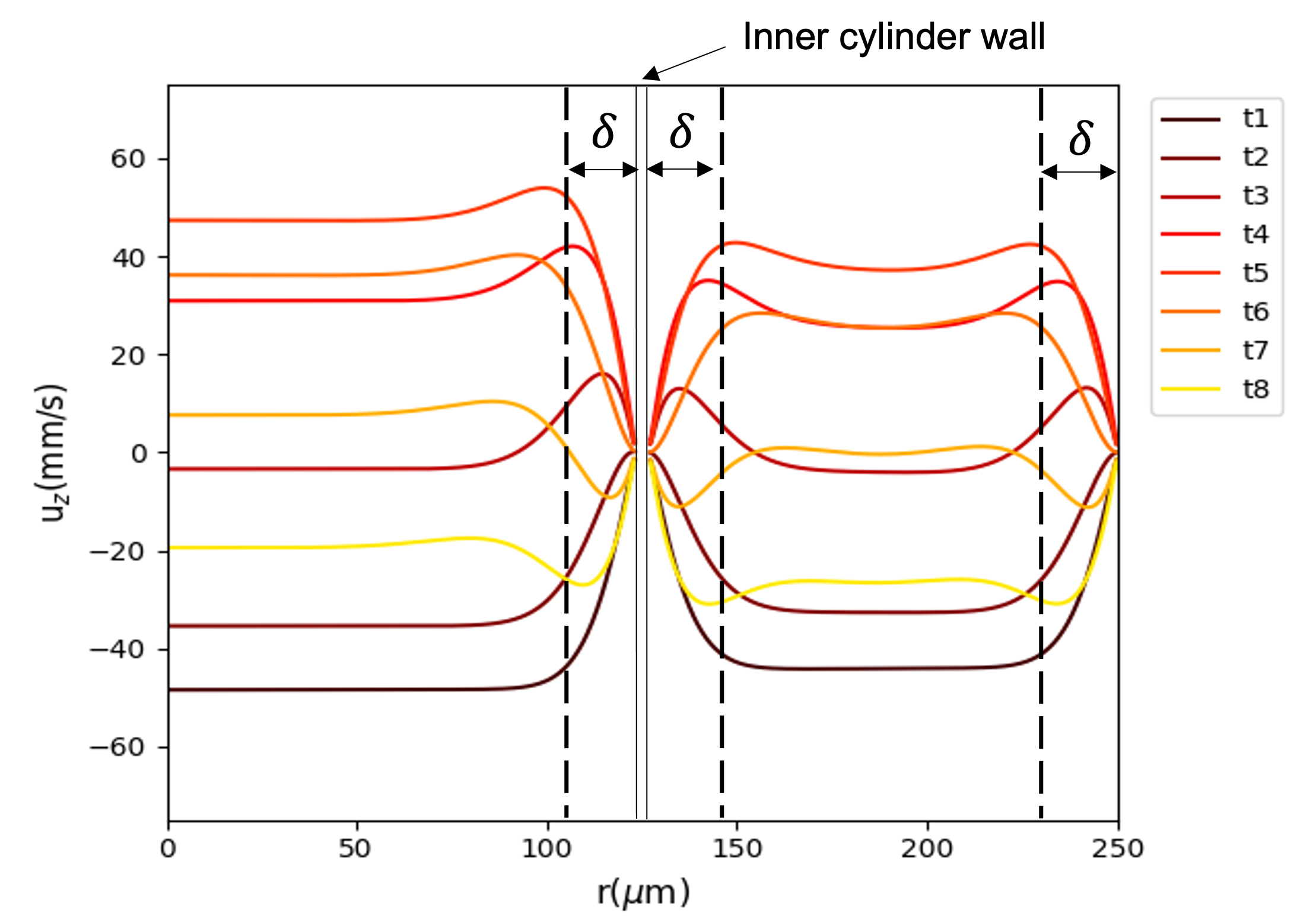}
	\caption{Time snapshots of velocity profiles generated by the meniscus oscillation inside a nozzle with a concentric cylinder ($n=1$) after a pressure pulse is applied on top. The cylindrical nozzle radius is $R=250\mu m$. From brown to yellow ($t_1-t_8$), the axial velocity $u_z$ is plotted in the increasing order in time every 0.1 ms. The velocity profiles are measures at $z=200\mu m$ above the nozzle outlet.}
	\label{fig:bl2}
\end{figure}

In Fig. \ref{fig:velocity} and Fig. \ref{fig:bl2} we plot the vertical velocity contours and  profiles inside the nozzle with concentric cylinders. They show the formation of thin boundary layers near the surface of the inner cylinder(s) and the standard cylinder wall. In Fig. \ref{fig:velocity}, we observe that, as the number of inner cylinders is increased, more layers of boundary layers contribute to increasing the energy dissipation in the liquid. The boundary layer thickness at the inner cylinder (Fig. \ref{fig:bl2}) is similar to that in the standard nozzle wall, $\delta \approx 20\mu m$, since the oscillation frequency stays nearly constant.

    \section{Discussion} \label{sec:6}

The nozzle geometry with equally spaced concentric cylinders is a successful example of dissipation control without introducing a cross-sectional area constriction. However, the concentric cylinders are a concept geometry and not a favorable geometry for fabrication. Practical nozzle design should be able to consider ease of manufacturing. 

\begin{figure}
	\vspace{-6pt}
	\centering
	\includegraphics[width=0.4\textwidth, keepaspectratio]{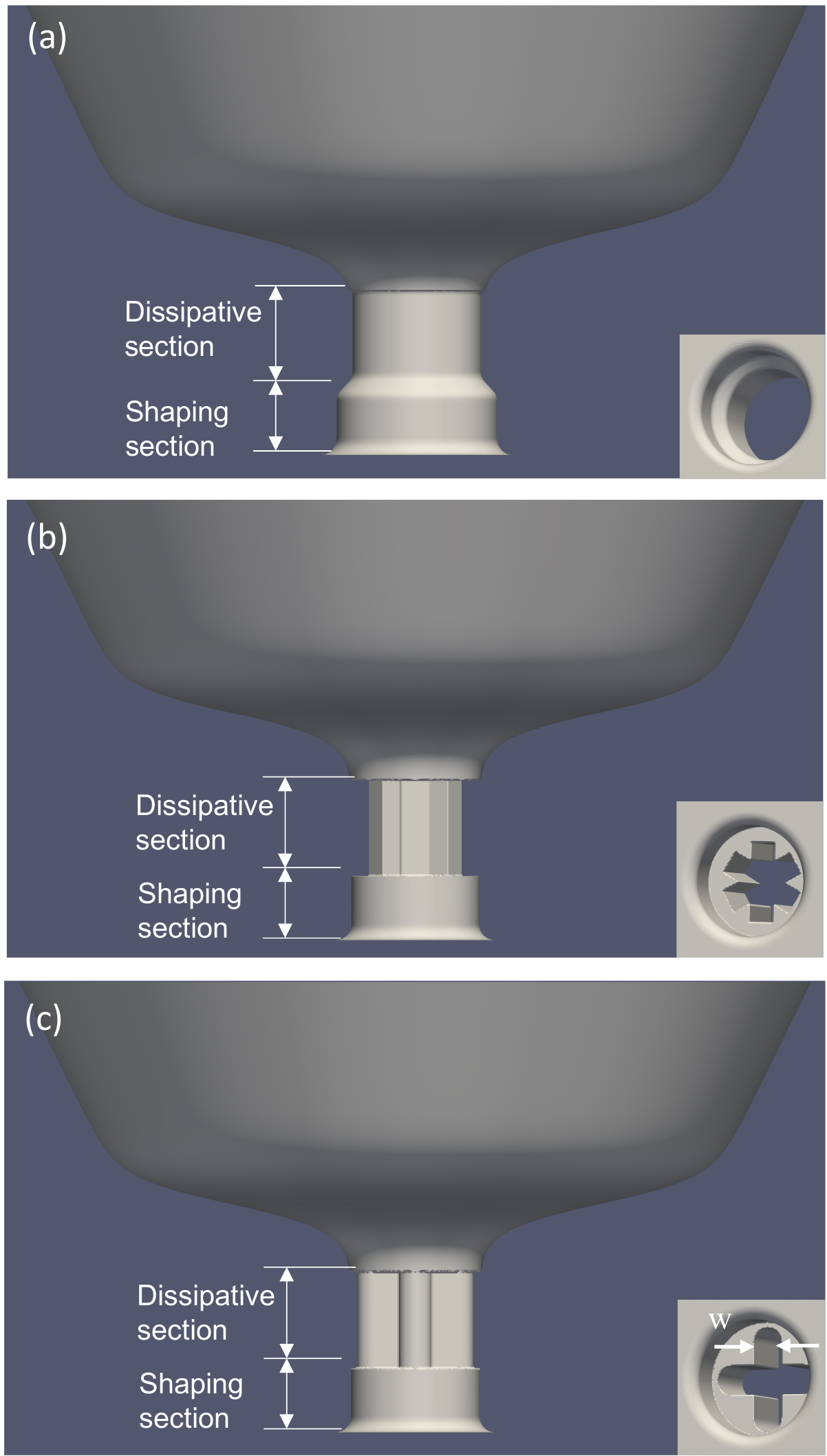}
	\caption{Example of nozzle designs with a dissipative section potentially simple to manufacture. The inset figures are cross-sectional views of the nozzle outlet. (a) A nozzle with a dissipative section with a constriction; (b) A nozzle with a dissipative section formed by an extruded star-shaped cross-section (sharp edges may be smoothed); (c) A nozzle with a dissipative section formed by an extruded cross-shaped cross-section. Here $w$ is the width of a branch of the cross.}
	\label{fig:manufacturing}
\end{figure}

In Fig. \ref{fig:manufacturing}, we have listed several examples of manufacturable nozzle designs inspired by the present analysis. The constricted nozzle (Fig. \ref{fig:manufacturing} (a)) can be easily fabricated using 3D printing or machine tools. A drawback of this geometry, when the constriction ratio  $R/R_i$ is large enough, is  the flow separation and formation of vortices at the intersection between {\it dissipative} and {\it shaping} section. This separation can lead to a very inhomogeneous velocity field in the shaping section that affects the formation of the droplet. For liquid metals, these vortices can easily form for values of $R/R_i$ that are not too far from $1$, albeit the droplet shape may not be affected at such low ratios. 

The other class of nozzle geometry has a cross-section extruded along the flow direction (e.g., the star- or cross-shaped  geometry shown in Fig. \ref{fig:manufacturing} (b) and (c)). These geometries are expected to promote viscous dissipation because both the cross-sectional area and the volume-to-surface area ratio are smaller than in the standard cylindrical nozzle. While flow separation can also occur in these geometries in the transition between the dissipative and shaping section, they better "distribute" the faster fluid leaving the dissipative section. This helps forming a more homogeneous velocity field in the ejected droplet, and hence to more regular jetting. This type of nozzle geometries can be  manufactured by drilling, or potentially 3D printing. 

Advancing the concept of the cross or the star cross-sections further, cross-sections with $n$ straight channels of width $w$ separated by the same angle stemming from the center of the cross-section can be considered (Fig. \ref{fig:manufacturing} (c)). This makes it possible to control the surface area $S$ while while keeping the volume $V$ roughly constant, by increasing $n$ and decreasing $w$.  The width $w$ should be chosen based on the estimation of the thickness of the boundary layer. When $w \lesssim 2\delta$, the velocity profile inside the branch will be close to parabolic. In this case, the scaling of viscous dissipation will be modified, since $\delta \sim w$, so that $\gamma_\text{bl}\sim (\nu/w) (S/V) \sim \nu/w^2$, assuming $w \sim \delta \ll R$. The value of $n$ should be chosen as a compromise between manufacturability, volume of the dissipative section, and droplet shape. 

    \section{Conclusion and Summary} \label{sec:7}

Drop-on-demand liquid metal jetting is an emerging technology that enables the creation of quality products and provides a safe printing process. To enable reliable high-speed printing, it is required to develop a nozzle that promotes fast decay of the meniscus oscillation after a droplet ejection. Design guidelines for high-speed jetting need a physical understanding of the energy dissipation mechanisms that controls the relaxation time of the fluid in the nozzle. Dimensional analysis reveals that the oscillatory flow inside the nozzle, driven by the meniscus motion, is in the inertial flow regime, and that it forms a thin boundary layer near the solid wall. In this paper, we propose physics-based design rules to speed up the decay of the meniscus fluctuations, based on scaling relations derived from the Navier-Stokes equation and the viscous boundary layer theory. These scaling  relations informs that nozzles with smaller diameters and nozzles that maximize the surface area to volume ratio have faster damping rates. These dimensional considerations need to be balanced by the need to obtain droplets of a given shape and size, so we focus on the design of nozzles with an outlet of a given cross sectional area. To demonstrate these ideas, we performed multiphase flow simulations for several nozzle designs. We used a class of idealized geometries that include multiple equally spaced concentric cylinders perpendicular to the nozzle meniscus movement. The simulation results showed that the damping rate increases as the surface area of the nozzle does, and displays the  linear growth with the number of concentric cylinders predicted by the scaling relations. Probing the vertical velocity profiles inside the nozzle confirmed the formation of a thin boundary layer near the wall. 

Finally, we discussed a class of manufacturable nozzle designs inspired by the discussion herein. To the best of our knowledge, our study is the first investigation of the damping rate in a liquid metal jetting printer nozzle using  oscillatory viscous boundary layer theory. Also, our study is the first demonstration of dissipation control by introducing design changes in the nozzle shape. We note that an accurate estimation of the meniscus damping in numerical simulations is challenging due to the multiple sources of dissipation, including the boundary layer, bulk, contact line, and numerical dissipation. A systematic  and rigorous analysis of the sources of viscous dissipation, and a demanding numerical convergence study, were crucial prerequisites for the results herein. 

This study also suggests a number of questions for future investigation. First, guidelines for the design of the shaping section are needed, in which its shape should promote the homogenization of the velocity field prior to droplet formation. This would minimize the likelihood of appearance of multiple droplets as a result of a single pressure pulse. Second, experimental validation of the ideas herein is pending. Finally, accurately computing the damping rate of non-axisymmetric nozzles (fully three-dimensional geometries) is very computationally demanding, and hence was left for future studies.

    \section{Numerical Methods and Validation} \label{sec:8}

We describe the numerical tools we used to discretize the governing equations and to perform the simulations of  multiphase flows shown here. All of them were computed with the \interfoam\;solver in the OpenFOAM version 2106. 

\subsection{Interface Capturing Scheme}

In \interfoam, the algebraic Volume-of-Fluid (VoF) interface tracking method is applied to track the dynamics of the gas-liquid interface. In a discretized domain with VOF, the fraction of volume of liquid, $\alpha$, can take a value between 0 and 1 near the interface. As a result, in numerical simulations the gas-liquid interface has a finite thickness,  instead of the sharp interface described as a Dirac delta function in Eq. (\ref{eq:NS}). Using the Continuum Surface Force model (CSF)\cite{Brackbill1992}, the surface tension force in the momentum Eq. (\ref{eq:NS}) is calculated as \eq{CSF}{\sigma \kappa\vec{n}\hat{\delta}(\vec{x}-\vec{x}_s) \approx\sigma \kappa \nabla \alpha.} Given the volume fraction of fluid, $\alpha(\vec{x},t)$, computed at each time $t$ and position $\vec {x}$, the viscosity and the density fields are updated as 
\eq{mu}{\mu_f(\vec{x},t)=\alpha(\vec{x},t)\mu + (1-\alpha(\vec{x},t))\mu_g,} and \eq{rho}{\rho_f(\vec{x},t)=\alpha(\vec{x},t)\rho + (1-\alpha(\vec{x},t))\rho_g,} and used to solve the N-S equation. We refer to Desphande et al. \cite{Deshpande2012} for a description of detailed algorithms and finite volume formulations to solve the N-S equation and the volume fraction field.

\subsection{Temporal and Spatial Discretization}

OpenFOAM employs a cell-centered finite volume discretization. The spatial domain is discretized with hexahedral meshes generated by the built-in meshing tool, {\sffamily blockMesh} and {\sffamily snappyHexMesh}. We chose the linear scheme for cell center-face interpolation, and the Gauss linear scheme for divergence, gradient, and Laplacian operators. 

The time discretization was performed via the explicit Euler time scheme. A variable time-step method was chosen to ensure numerical stability, and controlled by the time step restrictions based on a Courant-Freidich-Lewis(CFL) number derived  from the momentum equation and the surface tension\cite{Deshpande2012}. We enforced CFL numbers below 0.2 for both momentum and interface driven time-step restrictions.

The solution of the momentum equation was obtained by constructing a predicted velocity field and correcting it using the Pressure Implicit with Splitting of Operators (PISO) algorithm \cite{Issa1986}. For the Poisson equation  in this pressure correction, we used the preconditioned conjugate gradient scheme with the diagonal-based incomplete Cholesky preconditioner. 

We turned off the interface compression method in \interfoam\;by setting {\sffamily cAlpha}=0 in {\sffamily fvSolution}. Instead, we created extra refined meshes near the region of the interface to avoid numerical diffusion of the interface thickness. 

\subsection{Validation}

The \interfoam\;solver has been validated in previous studies \cite{Deshpande2012}, and it showed nice agreement in tracking interfaces for highly inertial flows and surface-tension-driven capillary flows. Standard tests showed that the CSF formulation in  \interfoam\;led to a proper  discrete balance between pressure and  surface tension \cite{Deshpande2012}. 

Although details are not presented here,  we independently performed validation studies to assess the quantitative predictive ability of the \interfoam\;solver on two-phase problems with oscillating liquid-gas interfaces. We evaluated extensive simulation results from \interfoam\;against reference solutions in canonical multiphase flow test cases in the high capillarity limit, obtained by analytical investigations \cite{Lamb1932,Case1956,Miles1998,Henderson1994} with matched experimental investigations \cite{Howell1999,Becker1991}, or numerical solutions obtained by other numerical methods on highly resolved grids (level set \cite{Herrmann2008} or phase field\cite{Mirjalili2019}). These canonical validation test cases involve damped oscillations of initially perturbed two-dimensional circular (cylinder) and spherical droplets \cite{Mirjalili2019,Herrmann2008}, of capillary waves in a periodic domain \cite{Herrmann2008}, and of the brimful cylindrical tank with water \cite{Howell1999, Case1956, Henderson1994}. These simulations were performed with material properties of either the water-air or the liquid metal-argon systems. We confirmed that the oscillation frequency and the damping rate on perturbed capillary flows calculated by the \interfoam\;solver showed nice agreement against the reference solutions within $\sim10\%$ errors. We note that the numerical dissipation of \interfoam, especially on some coarse grids, contributed to the overestimation of the damping rate, and we acknowledged that the grid convergence study was a critical component of the validation. We tested three-step grid refinements and demonstrated the grid convergence on both damping rate and oscillation frequency.

    \nocite{*}
    \bibliography{nozzleDesign}% Produces the bibliography via BibTeX.

\providecommand{\noopsort}[1]{}\providecommand{\singleletter}[1]{#1}%
\begin{thebibliography}{23}%
\makeatletter
\providecommand \@ifxundefined [1]{%
 \@ifx{#1\undefined}
}%
\providecommand \@ifnum [1]{%
 \ifnum #1\expandafter \@firstoftwo
 \else \expandafter \@secondoftwo
 \fi
}%
\providecommand \@ifx [1]{%
 \ifx #1\expandafter \@firstoftwo
 \else \expandafter \@secondoftwo
 \fi
}%
\providecommand \natexlab [1]{#1}%
\providecommand \enquote  [1]{``#1''}%
\providecommand \bibnamefont  [1]{#1}%
\providecommand \bibfnamefont [1]{#1}%
\providecommand \citenamefont [1]{#1}%
\providecommand \href@noop [0]{\@secondoftwo}%
\providecommand \href [0]{\begingroup \@sanitize@url \@href}%
\providecommand \@href[1]{\@@startlink{#1}\@@href}%
\providecommand \@@href[1]{\endgroup#1\@@endlink}%
\providecommand \@sanitize@url [0]{\catcode `\\12\catcode `\$12\catcode
  `\&12\catcode `\#12\catcode `\^12\catcode `\_12\catcode `\%12\relax}%
\providecommand \@@startlink[1]{}%
\providecommand \@@endlink[0]{}%
\providecommand \url  [0]{\begingroup\@sanitize@url \@url }%
\providecommand \@url [1]{\endgroup\@href {#1}{\urlprefix }}%
\providecommand \urlprefix  [0]{URL }%
\providecommand \Eprint [0]{\href }%
\providecommand \doibase [0]{http://dx.doi.org/}%
\providecommand \selectlanguage [0]{\@gobble}%
\providecommand \bibinfo  [0]{\@secondoftwo}%
\providecommand \bibfield  [0]{\@secondoftwo}%
\providecommand \translation [1]{[#1]}%
\providecommand \BibitemOpen [0]{}%
\providecommand \bibitemStop [0]{}%
\providecommand \bibitemNoStop [0]{.\EOS\space}%
\providecommand \EOS [0]{\spacefactor3000\relax}%
\providecommand \BibitemShut  [1]{\csname bibitem#1\endcsname}%
\let\auto@bib@innerbib\@empty
%</preamble>
\bibitem [{\citenamefont {Sukhotskiy}\ \emph {et~al.}(2018)\citenamefont
  {Sukhotskiy}, \citenamefont {Vishnoi}, \citenamefont {Karampelas},
  \citenamefont {Vader}, \citenamefont {Vader},\ and\ \citenamefont
  {Furlani}}]{Sukhotskiy2018}%
  \BibitemOpen
  \bibfield  {author} {\bibinfo {author} {\bibfnamefont {V.}~\bibnamefont
  {Sukhotskiy}}, \bibinfo {author} {\bibfnamefont {P.}~\bibnamefont {Vishnoi}},
  \bibinfo {author} {\bibfnamefont {I.~H.}\ \bibnamefont {Karampelas}},
  \bibinfo {author} {\bibfnamefont {S.}~\bibnamefont {Vader}}, \bibinfo
  {author} {\bibfnamefont {Z.}~\bibnamefont {Vader}}, \ and\ \bibinfo {author}
  {\bibfnamefont {E.~P.}\ \bibnamefont {Furlani}},\ }\bibfield  {title}
  {\enquote {\bibinfo {title} {Magnetohydrodynamic drop-on-demand liquid metal
  additive manufacturing: System overview and modelling},}\ }\href@noop {}
  {\bibfield  {journal} {\bibinfo  {journal} {Proceedings of the 5th
  International Conference of Fluid Flow, Heat and Mass Transfer}\ ,\ \bibinfo
  {pages} {1--6}} (\bibinfo {year} {2018})}\BibitemShut {NoStop}%
\bibitem [{\citenamefont {Simonelli}\ \emph {et~al.}(2019)\citenamefont
  {Simonelli}, \citenamefont {Aboulkhair}, \citenamefont {Rasa}, \citenamefont
  {East}, \citenamefont {Tuck}, \citenamefont {Wildman}, \citenamefont
  {Salomons},\ and\ \citenamefont {Hague}}]{Simonelli2019}%
  \BibitemOpen
  \bibfield  {author} {\bibinfo {author} {\bibfnamefont {M.}~\bibnamefont
  {Simonelli}}, \bibinfo {author} {\bibfnamefont {N.}~\bibnamefont
  {Aboulkhair}}, \bibinfo {author} {\bibfnamefont {M.}~\bibnamefont {Rasa}},
  \bibinfo {author} {\bibfnamefont {M.}~\bibnamefont {East}}, \bibinfo {author}
  {\bibfnamefont {C.}~\bibnamefont {Tuck}}, \bibinfo {author} {\bibfnamefont
  {R.}~\bibnamefont {Wildman}}, \bibinfo {author} {\bibfnamefont
  {O.}~\bibnamefont {Salomons}}, \ and\ \bibinfo {author} {\bibfnamefont
  {R.}~\bibnamefont {Hague}},\ }\bibfield  {title} {\enquote {\bibinfo {title}
  {Towards digital metal additive manufacturing via high-temperature
  drop-on-demand jetting},}\ }\href@noop {} {\bibfield  {journal} {\bibinfo
  {journal} {Additive Manufacturing}\ }\textbf {\bibinfo {volume} {30}},\
  \bibinfo {pages} {1--9} (\bibinfo {year} {2019})}\BibitemShut {NoStop}%
\bibitem [{\citenamefont {Luo}\ \emph {et~al.}(2012)\citenamefont {Luo},
  \citenamefont {hua Qi}, \citenamefont {ming Zhou},\ and\ \citenamefont
  {Xiang-hui Houc}}]{Luo2012}%
  \BibitemOpen
  \bibfield  {author} {\bibinfo {author} {\bibfnamefont {J.}~\bibnamefont
  {Luo}}, \bibinfo {author} {\bibfnamefont {L.}~\bibnamefont {hua Qi}},
  \bibinfo {author} {\bibfnamefont {J.}~\bibnamefont {ming Zhou}}, \ and\
  \bibinfo {author} {\bibfnamefont {H.-j.~L.}\ \bibnamefont {Xiang-hui Houc}},\
  }\bibfield  {title} {\enquote {\bibinfo {title} {Modeling and
  characterization of metal droplets generation by using a pneumatic
  drop-on-demand generator},}\ }\href@noop {} {\bibfield  {journal} {\bibinfo
  {journal} {Journal of Materials Processing Technology}\ }\textbf {\bibinfo
  {volume} {212}},\ \bibinfo {pages} {718--726} (\bibinfo {year}
  {2012})}\BibitemShut {NoStop}%
\bibitem [{\citenamefont {Howell}\ \emph {et~al.}(1999)\citenamefont {Howell},
  \citenamefont {Burhow}, \citenamefont {Heath}, \citenamefont {McKenna},
  \citenamefont {Hwang},\ and\ \citenamefont {Schatz}}]{Howell1999}%
  \BibitemOpen
  \bibfield  {author} {\bibinfo {author} {\bibfnamefont {D.~R.}\ \bibnamefont
  {Howell}}, \bibinfo {author} {\bibfnamefont {B.}~\bibnamefont {Burhow}},
  \bibinfo {author} {\bibfnamefont {T.}~\bibnamefont {Heath}}, \bibinfo
  {author} {\bibfnamefont {C.}~\bibnamefont {McKenna}}, \bibinfo {author}
  {\bibfnamefont {W.}~\bibnamefont {Hwang}}, \ and\ \bibinfo {author}
  {\bibfnamefont {M.~F.}\ \bibnamefont {Schatz}},\ }\bibfield  {title}
  {\enquote {\bibinfo {title} {Measurements of surface-wave damping in a
  container},}\ }\href@noop {} {\bibfield  {journal} {\bibinfo  {journal}
  {Physics of Fluids}\ }\textbf {\bibinfo {volume} {12}},\ \bibinfo {pages}
  {322--326} (\bibinfo {year} {1999})}\BibitemShut {NoStop}%
\bibitem [{\citenamefont {Ting}\ and\ \citenamefont {Perlin}(1995)}]{Ting1995}%
  \BibitemOpen
  \bibfield  {author} {\bibinfo {author} {\bibfnamefont {C.-L.}\ \bibnamefont
  {Ting}}\ and\ \bibinfo {author} {\bibfnamefont {M.}~\bibnamefont {Perlin}},\
  }\bibfield  {title} {\enquote {\bibinfo {title} {Boundary conditions in the
  vicinity of the contact line at a vertically oscillating upright plate : an
  experimental investigation},}\ }\href@noop {} {\bibfield  {journal} {\bibinfo
   {journal} {Journal of Fluid Mechanics}\ }\textbf {\bibinfo {volume} {295}},\
  \bibinfo {pages} {263--300} (\bibinfo {year} {1995})}\BibitemShut {NoStop}%
\bibitem [{\citenamefont {Stachewicz}\ \emph {et~al.}(2009)\citenamefont
  {Stachewicz}, \citenamefont {Dijksman}, \citenamefont {Burdinski},
  \citenamefont {Yurteri},\ and\ \citenamefont
  {Marijnissen}}]{Stacheswicz2009}%
  \BibitemOpen
  \bibfield  {author} {\bibinfo {author} {\bibfnamefont {U.}~\bibnamefont
  {Stachewicz}}, \bibinfo {author} {\bibfnamefont {J.~F.}\ \bibnamefont
  {Dijksman}}, \bibinfo {author} {\bibfnamefont {D.}~\bibnamefont {Burdinski}},
  \bibinfo {author} {\bibfnamefont {C.~U.}\ \bibnamefont {Yurteri}}, \ and\
  \bibinfo {author} {\bibfnamefont {J.~C.~M.}\ \bibnamefont {Marijnissen}},\
  }\bibfield  {title} {\enquote {\bibinfo {title} {Relaxation times in single
  event electrospraying controlled by nozzle front surface modification},}\
  }\href@noop {} {\bibfield  {journal} {\bibinfo  {journal} {Langmuir}\
  }\textbf {\bibinfo {volume} {25}},\ \bibinfo {pages} {2540--2549} (\bibinfo
  {year} {2009})}\BibitemShut {NoStop}%
\bibitem [{\citenamefont {Seo}, \citenamefont {Garcia-Mayoral},\ and\
  \citenamefont {Mani}(2015)}]{Seo2015}%
  \BibitemOpen
  \bibfield  {author} {\bibinfo {author} {\bibfnamefont {J.}~\bibnamefont
  {Seo}}, \bibinfo {author} {\bibfnamefont {R.}~\bibnamefont {Garcia-Mayoral}},
  \ and\ \bibinfo {author} {\bibfnamefont {A.}~\bibnamefont {Mani}},\
  }\bibfield  {title} {\enquote {\bibinfo {title} {Pressure fluctuations and
  interfacial robustness in turbulent flows over superhydrophobic surfaces},}\
  }\href@noop {} {\bibfield  {journal} {\bibinfo  {journal} {Journal of Fluid
  Mechanics}\ }\textbf {\bibinfo {volume} {783}},\ \bibinfo {pages} {448--473}
  (\bibinfo {year} {2015})}\BibitemShut {NoStop}%
\bibitem [{\citenamefont {Snoeijer}\ and\ \citenamefont
  {Andreotti}(2013)}]{Snoeijer2013}%
  \BibitemOpen
  \bibfield  {author} {\bibinfo {author} {\bibfnamefont {J.~H.}\ \bibnamefont
  {Snoeijer}}\ and\ \bibinfo {author} {\bibfnamefont {B.}~\bibnamefont
  {Andreotti}},\ }\bibfield  {title} {\enquote {\bibinfo {title} {Moving
  contact lines: Scales, regimes, and dynamical transitions},}\ }\href@noop {}
  {\bibfield  {journal} {\bibinfo  {journal} {Annu. Rev. Fluid Mech}\ }\textbf
  {\bibinfo {volume} {45}},\ \bibinfo {pages} {269--292} (\bibinfo {year}
  {2013})}\BibitemShut {NoStop}%
\bibitem [{\citenamefont {Assael}\ \emph {et~al.}(2006)\citenamefont {Assael},
  \citenamefont {Kakosimos}, \citenamefont {Banish}, \citenamefont {Brillo},
  \citenamefont {Egry}, \citenamefont {Brooks}, \citenamefont {Quested},
  \citenamefont {Mills}, \citenamefont {Nagashima}, \citenamefont {Sato},\ and\
  \citenamefont {Wakeham}}]{Assael2006}%
  \BibitemOpen
  \bibfield  {author} {\bibinfo {author} {\bibfnamefont {M.~J.}\ \bibnamefont
  {Assael}}, \bibinfo {author} {\bibfnamefont {K.}~\bibnamefont {Kakosimos}},
  \bibinfo {author} {\bibfnamefont {R.~M.}\ \bibnamefont {Banish}}, \bibinfo
  {author} {\bibfnamefont {J.}~\bibnamefont {Brillo}}, \bibinfo {author}
  {\bibfnamefont {I.}~\bibnamefont {Egry}}, \bibinfo {author} {\bibfnamefont
  {R.}~\bibnamefont {Brooks}}, \bibinfo {author} {\bibfnamefont {P.~N.}\
  \bibnamefont {Quested}}, \bibinfo {author} {\bibfnamefont {K.~C.}\
  \bibnamefont {Mills}}, \bibinfo {author} {\bibfnamefont {A.}~\bibnamefont
  {Nagashima}}, \bibinfo {author} {\bibfnamefont {Y.}~\bibnamefont {Sato}}, \
  and\ \bibinfo {author} {\bibfnamefont {W.~A.}\ \bibnamefont {Wakeham}},\
  }\bibfield  {title} {\enquote {\bibinfo {title} {Reference data for the
  density and viscosity of liquid aluminum and liquid iron},}\ }\href@noop {}
  {\bibfield  {journal} {\bibinfo  {journal} {J. Phys. Chem. Ref. Data}\
  }\textbf {\bibinfo {volume} {35}},\ \bibinfo {pages} {285--2006} (\bibinfo
  {year} {2006})}\BibitemShut {NoStop}%
\bibitem [{\citenamefont {Molina}\ \emph {et~al.}(2007)\citenamefont {Molina},
  \citenamefont {Voytovych}, \citenamefont {Louis},\ and\ \citenamefont
  {Eustathopoulos}}]{Molina2007}%
  \BibitemOpen
  \bibfield  {author} {\bibinfo {author} {\bibfnamefont {J.}~\bibnamefont
  {Molina}}, \bibinfo {author} {\bibfnamefont {R.}~\bibnamefont {Voytovych}},
  \bibinfo {author} {\bibfnamefont {E.}~\bibnamefont {Louis}}, \ and\ \bibinfo
  {author} {\bibfnamefont {N.}~\bibnamefont {Eustathopoulos}},\ }\bibfield
  {title} {\enquote {\bibinfo {title} {The surface tension of liquid aluminium
  in high vacuum: The role of surface condition},}\ }\href@noop {} {\bibfield
  {journal} {\bibinfo  {journal} {International Journal of Adhesion and
  Adhesives}\ }\textbf {\bibinfo {volume} {27}},\ \bibinfo {pages} {394--401}
  (\bibinfo {year} {2007})}\BibitemShut {NoStop}%
\bibitem [{\citenamefont {Landau}\ and\ \citenamefont
  {Lifshitz}(1958)}]{Landau1958}%
  \BibitemOpen
  \bibfield  {author} {\bibinfo {author} {\bibfnamefont {L.}~\bibnamefont
  {Landau}}\ and\ \bibinfo {author} {\bibfnamefont {E.~M.}\ \bibnamefont
  {Lifshitz}},\ }\href@noop {} {\emph {\bibinfo {title} {Fluid Mechanics}}}\
  (\bibinfo  {publisher} {Pergamon Press},\ \bibinfo {year} {1958})\BibitemShut
  {NoStop}%
\bibitem [{\citenamefont {Lamb}(1932)}]{Lamb1932}%
  \BibitemOpen
  \bibfield  {author} {\bibinfo {author} {\bibfnamefont {H.}~\bibnamefont
  {Lamb}},\ }\href@noop {} {\emph {\bibinfo {title} {Hydrodynamics}}}\
  (\bibinfo  {publisher} {Cambridge University Press.},\ \bibinfo {year}
  {1932})\BibitemShut {NoStop}%
\bibitem [{\citenamefont {Henderson}\ and\ \citenamefont
  {Miles}(1994)}]{Henderson1994}%
  \BibitemOpen
  \bibfield  {author} {\bibinfo {author} {\bibfnamefont {D.~M.}\ \bibnamefont
  {Henderson}}\ and\ \bibinfo {author} {\bibfnamefont {J.~W.}\ \bibnamefont
  {Miles}},\ }\bibfield  {title} {\enquote {\bibinfo {title} {Surface-wave
  damping in a circular cylinder with a fixed contact line},}\ }\href@noop {}
  {\bibfield  {journal} {\bibinfo  {journal} {Journal of Fluid Mechanics}\
  }\textbf {\bibinfo {volume} {275}},\ \bibinfo {pages} {285--299} (\bibinfo
  {year} {1994})}\BibitemShut {NoStop}%
\bibitem [{\citenamefont {Case}\ and\ \citenamefont
  {Parkinson}(1956)}]{Case1956}%
  \BibitemOpen
  \bibfield  {author} {\bibinfo {author} {\bibfnamefont {K.~M.}\ \bibnamefont
  {Case}}\ and\ \bibinfo {author} {\bibfnamefont {W.~C.}\ \bibnamefont
  {Parkinson}},\ }\bibfield  {title} {\enquote {\bibinfo {title} {Damping of
  surface waves in an incompressible liquid},}\ }\href@noop {} {\bibfield
  {journal} {\bibinfo  {journal} {Journal of Fluid Mechanics}\ }\textbf
  {\bibinfo {volume} {2}},\ \bibinfo {pages} {172--184} (\bibinfo {year}
  {1956})}\BibitemShut {NoStop}%
\bibitem [{Note1()}]{Note1}%
  \BibitemOpen
  \bibinfo {note} {Here $n=1$ is the standard nozzle, i.e. with zero internal
  concentric cylinders.}\BibitemShut {Stop}%
\bibitem [{\citenamefont {Brackbill}, \citenamefont {Kothe},\ and\
  \citenamefont {Zemach}(1992)}]{Brackbill1992}%
  \BibitemOpen
  \bibfield  {author} {\bibinfo {author} {\bibfnamefont {J.~U.}\ \bibnamefont
  {Brackbill}}, \bibinfo {author} {\bibfnamefont {D.~B.}\ \bibnamefont
  {Kothe}}, \ and\ \bibinfo {author} {\bibfnamefont {C.}~\bibnamefont
  {Zemach}},\ }\bibfield  {title} {\enquote {\bibinfo {title} {A continuum
  method for modeling surface tension},}\ }\href@noop {} {\bibfield  {journal}
  {\bibinfo  {journal} {J. Comput. Phys.}\ }\textbf {\bibinfo {volume} {100}},\
  \bibinfo {pages} {335--354} (\bibinfo {year} {1992})}\BibitemShut {NoStop}%
\bibitem [{\citenamefont {Deshpande}, \citenamefont {Anumolu},\ and\
  \citenamefont {Trujillo}(2012)}]{Deshpande2012}%
  \BibitemOpen
  \bibfield  {author} {\bibinfo {author} {\bibfnamefont {S.~S.}\ \bibnamefont
  {Deshpande}}, \bibinfo {author} {\bibfnamefont {L.}~\bibnamefont {Anumolu}},
  \ and\ \bibinfo {author} {\bibfnamefont {M.~F.}\ \bibnamefont {Trujillo}},\
  }\bibfield  {title} {\enquote {\bibinfo {title} {Evaluating the performance
  of the two-phase flow solver interfoam},}\ }\href@noop {} {\bibfield
  {journal} {\bibinfo  {journal} {Computational Science and Discovery}\
  }\textbf {\bibinfo {volume} {5}},\ \bibinfo {pages} {014016} (\bibinfo {year}
  {2012})}\BibitemShut {NoStop}%
\bibitem [{\citenamefont {Issa}(1986)}]{Issa1986}%
  \BibitemOpen
  \bibfield  {author} {\bibinfo {author} {\bibfnamefont {R.~I.}\ \bibnamefont
  {Issa}},\ }\bibfield  {title} {\enquote {\bibinfo {title} {Solution of the
  implicitly discretised fluid flow equations by operator splitting},}\
  }\href@noop {} {\bibfield  {journal} {\bibinfo  {journal} {J. Comput. Phys.}\
  }\textbf {\bibinfo {volume} {62}},\ \bibinfo {pages} {40--65} (\bibinfo
  {year} {1986})}\BibitemShut {NoStop}%
\bibitem [{\citenamefont {Miles}\ and\ \citenamefont
  {Henderson}(1998)}]{Miles1998}%
  \BibitemOpen
  \bibfield  {author} {\bibinfo {author} {\bibfnamefont {J.~W.}\ \bibnamefont
  {Miles}}\ and\ \bibinfo {author} {\bibfnamefont {D.~M.}\ \bibnamefont
  {Henderson}},\ }\bibfield  {title} {\enquote {\bibinfo {title} {A note on
  interior vs. boundary-layer damping of surface waves in a circular
  cylinder},}\ }\href@noop {} {\bibfield  {journal} {\bibinfo  {journal}
  {Journal of Fluid Mechanics}\ }\textbf {\bibinfo {volume} {364}},\ \bibinfo
  {pages} {319--323} (\bibinfo {year} {1998})}\BibitemShut {NoStop}%
\bibitem [{\citenamefont {Becker}, \citenamefont {Hiller},\ and\ \citenamefont
  {Kowaleswski}(1991)}]{Becker1991}%
  \BibitemOpen
  \bibfield  {author} {\bibinfo {author} {\bibfnamefont {E.}~\bibnamefont
  {Becker}}, \bibinfo {author} {\bibfnamefont {W.~J.}\ \bibnamefont {Hiller}},
  \ and\ \bibinfo {author} {\bibfnamefont {T.~A.}\ \bibnamefont
  {Kowaleswski}},\ }\bibfield  {title} {\enquote {\bibinfo {title}
  {Experimental and theoretical investigation of large amplitude oscillations
  of liquid droplets},}\ }\href@noop {} {\bibfield  {journal} {\bibinfo
  {journal} {Journal of Fluid Mechanics}\ }\textbf {\bibinfo {volume} {231}},\
  \bibinfo {pages} {189--210} (\bibinfo {year} {1991})}\BibitemShut {NoStop}%
\bibitem [{\citenamefont {Herrmann}(2008)}]{Herrmann2008}%
  \BibitemOpen
  \bibfield  {author} {\bibinfo {author} {\bibfnamefont {M.}~\bibnamefont
  {Herrmann}},\ }\bibfield  {title} {\enquote {\bibinfo {title} {A balanced
  force refined level set grid method for two-phase flows on unstructured flow
  solver grids},}\ }\href@noop {} {\bibfield  {journal} {\bibinfo  {journal}
  {Journal of Computational Physics}\ }\textbf {\bibinfo {volume} {227}},\
  \bibinfo {pages} {2674--2706} (\bibinfo {year} {2008})}\BibitemShut {NoStop}%
\bibitem [{\citenamefont {Mirjalili}, \citenamefont {Ivey},\ and\ \citenamefont
  {Mani}(2019)}]{Mirjalili2019}%
  \BibitemOpen
  \bibfield  {author} {\bibinfo {author} {\bibfnamefont {S.}~\bibnamefont
  {Mirjalili}}, \bibinfo {author} {\bibfnamefont {C.~B.}\ \bibnamefont {Ivey}},
  \ and\ \bibinfo {author} {\bibfnamefont {A.}~\bibnamefont {Mani}},\
  }\bibfield  {title} {\enquote {\bibinfo {title} {Comparison between the
  diffuse interface and volume of fluid methods for simulating two-phase
  flows},}\ }\href@noop {} {\bibfield  {journal} {\bibinfo  {journal}
  {International Journal of Multiphase Flow}\ }\textbf {\bibinfo {volume}
  {116}},\ \bibinfo {pages} {221--238} (\bibinfo {year} {2019})}\BibitemShut
  {NoStop}%
\bibitem [{\citenamefont {Prosperetti}(1981)}]{Prosperetti1981}%
  \BibitemOpen
  \bibfield  {author} {\bibinfo {author} {\bibfnamefont {A.}~\bibnamefont
  {Prosperetti}},\ }\bibfield  {title} {\enquote {\bibinfo {title} {Motion of
  two superposed viscous fluids},}\ }\href@noop {} {\bibfield  {journal}
  {\bibinfo  {journal} {Physics of Fluids}\ }\textbf {\bibinfo {volume} {24}},\
  \bibinfo {pages} {1217--1223} (\bibinfo {year} {1981})}\BibitemShut {NoStop}%
\end{thebibliography}%

\end{document}